\documentclass[conference]{IEEEtran}
\IEEEoverridecommandlockouts
\usepackage{cite}
\usepackage{amsmath,amssymb,amsfonts}
\usepackage{algorithmic}
\usepackage{graphicx}
\usepackage{textcomp}
\usepackage{xcolor}

\def\BibTeX{{\rm B\kern-.05em{\sc i\kern-.025em b}\kern-.08em
    T\kern-.1667em\lower.7ex\hbox{E}\kern-.125emX}}
\begin{document}

\title{Practical Quantum Search by Variational Quantum\\ Eigensolver on Noisy Intermediate-scale\\ Quantum Hardware
}

\author{


\IEEEauthorblockN{
    Chen-Yu Liu \IEEEauthorrefmark{1}\IEEEauthorrefmark{2}
}
\IEEEauthorblockA{\IEEEauthorrefmark{1} Hon Hai Research Institute, Taipei, Taiwan}
\IEEEauthorblockA{\IEEEauthorrefmark{2} Graduate Institute of Applied Physics, National Taiwan University, Taipei, Taiwan}
\IEEEauthorblockA{Email:  d10245003@g.ntu.edu.tw}
}

\maketitle

\begin{abstract}
Grover search is a renowned quantum search algorithm that leverages quantum superposition to find a marked item with quadratic speedup. However, when implemented on Noisy Intermediate-scale Quantum (NISQ) hardware, the required repeated iterations of the oracle and diffusion operators increase exponentially with the number of qubits, resulting in significant noise accumulation. To address this, we propose a hybrid quantum-classical architecture that replaces quantum iterations with updates from a classical optimizer. This optimizer minimizes the expectation value of an oracle Hamiltonian with respect to a parameterized quantum state representing the target bit string. Our parameterized quantum circuit is much shallower than Grover search circuit, and we found that it outperforms Grover search on noisy simulators and NISQ hardware. When the number of qubits is greater than 5, our approach still maintains usable success probability, while the success probability of Grover search is at the same level as random guessing.

\end{abstract}


\section{Introduction}

Quantum computing leverages the principles of quantum mechanics to perform calculations and solve problems~\cite{qc1, qc2}. It has the potential to solve certain problems much faster than classical computers, making it a promising technology for a wide range of applications. Before we enter the era of fault-tolerant quantum computing, in which quantum algorithms can provide advantages over classical computers, we are currently in the Noisy Intermediate-scale Quantum (NISQ) era ~\cite{qc2, nisq1}. In this era, the errors inherent in the quantum system limit the number of consecutive quantum operations that can be performed, which is also known as the circuit depth. 

NISQ algorithms and methods ~\cite{qc2, nisq1, nisq2} are proposed to face the challenge of the error and noise provided by the quantum hardware, such as Variational Quantum Eigensolver (VQE) ~\cite{vqe0, vqe1, vqe2, vqe3} and Quantum Error Mitigation (QEM) ~\cite{qem1, qem2, qem3, qem4}. VQE utilize a hybrid quantum-classical computational architecture to perform the expectation value optimization of the desired operator, which have wide range of applications in optimization problems ~\cite{vqe2, vqe4} and chemistry problems ~\cite{vqe1, vqe3}. With the limitation of the number of qubits in NISQ era, several methods to reduce the required number of qubits for the VQE are also proposed ~\cite{clustervqe1, qaoainqaoa1, qbeff1, hgaip}. 

Among all kinds of the quantum algorithms, Grover search provides a quadratic speedup over classical algorithms for searching an unstructured database. Specifically, Grover search can search an unsorted database of N items in $O(\sqrt{N})$ time, compared to the $O(N)$ time required by classical algorithms, where $N$ is the number of items in the database. This algorithm has been widely studied and has been shown to have numerous applications in quantum computing, such as database search ~\cite{gs1, gs2}, and quantum machine learning~\cite{gsml1, gsml2}.

Quantum machine learning (QML) is a novel approach that leverages the potential of quantum computers to facilitate the training and execution of machine learning algorithms \cite{qml1, qml2, qml3}. QML algorithms exploit the unique features of quantum computers such as superposition and interference to perform machine learning tasks more efficiently. The applications of QML are widespread, including image and speech recognition \cite{qml2}, natural language processing \cite{qnlp1}, recommendation systems \cite{qrs1, qrs2, qirs1, fsqc1}, and generative learning \cite{qgan1, qgan2, qgan3, qgan4, qgan5}. Despite the potential benefits of QML, it is an emerging field and several challenges need to be addressed before it can be widely adopted. These include the learnability \cite{qnnlearn1, qnnlearn2, qnnlearn3, qnnlearn4} and trainability \cite{qnntrain1, qnntrain2, qnntrain3, qnntrain4, qnntrain5} of QML models. In summary, QML offers significant promise as a powerful tool for analyzing complex data and has the potential to revolutionize many fields.

Implementations of Grover search have been made in several real quantum computer systems ~\cite{nisqgrover1, nisqgrover2, nisqgrover5, nisqgrover6, nisqgrover7}, with up to $5$-qubit ($N = 2^5$). One of the main challenges of implementing Grover search on NISQ hardware is the presence of noise and errors in the quantum gates, which can degrade the performance of the algorithm and lead to incorrect results. Additionally, NISQ hardware typically has a limited number of qubits, limiting the size of the search space that can be effectively searched using Grover Search. Furthermore, the iterative nature of the Grover iteration presents additional challenges in implementing the algorithm on NISQ hardware, such as the accumulated error and the requirement for longer quantum state coherence time.

Several approaches have been proposed to address the challenge of implementing Grover search on NISQ hardware, such as the quantum partial search algorithm~\cite{qps1, qps2}, the hybrid search algorithm~\cite{nisqgrover1, nisqgrover2}, the replacement of the global diffusion operator with local diffusion operators~\cite{nisqgrover2, nisqgrover3, nisqgrover4, nisqgrover5}, and the divide-and-conquer strategy~\cite{nisqgrover2}. The quantum partial search algorithm searches only a part of the target bit string, sacrificing speed for accuracy. The hybrid quantum search algorithm combines the use of local diffusion operators, which only act on a subset of the given database, with random guessing to obtain the full target string. The replacement of the global diffusion operator with local diffusion operators optimizes the depth of the circuit, making it perform better on NISQ hardware compared to the original approach. The divide-and-conquer strategy uses a multi-stage method to find part of the bit string and then renormalizes the search space to find the remaining target string.


The divided-and-conquer strategy was found to be the most effective of the strategies implemented on real hardware, as demonstrated by its success probability of 0.49 with 5 qubits on the Honeywell quantum computer with ion trap architecture~\cite{nisqgrover1}. However, the required circuit depth for these approaches is limited by the iterative nature of the Grover search, which scales as $O(\sqrt{N}) = O(2^{\frac{n}{2}})$, making it difficult to scale them up. Therefore, a scalable approach that can run on current NISQ hardware is essential for practical purposes in quantum search problems.


In this work, we reformulate a search problem as an optimization problem by incorporating an Oracle Hamiltonian and using the VQE to minimize the corresponding eigenvalue. By applying the VQE, we are able to achieve a quantum circuit depth that scales linearly with the length of the target bit string, in contrast to the exponential growth of the original Grover search. Our method therefore provides a scalable and practical solution on current NISQ hardware, representing a promising step towards achieving quantum advantage.


\subsection{Main Results} 
Our main results of using VQE as a hybrid quantum-classical search algorithm in both theoretical and experimental are summarized as follows:
\begin{itemize}
    \item Search task as an optimization task: For a task to search for a target bit string, an Oracle Hamiltonian is constructed, whose expectation value with respect to the parameterized quantum state provided by the VQE is the desired objective to be minimized, and the corresponding ground state is related to the target bit string. 

    \item Success probability: Our method achieve the highest success probability ever obtained from the real quantum computer to date, which is about 0.6 when $n = 5$ on 7-qubit ibm\_oslo and up to $\sim 0.4$ when $n = 9$ on 27-qubit ibm\_mumbai.
    
    
    \item Circuit depth reduction: In comparison to the Grover search, the depth of the quantum circuit required by the VQE search is reduced from $O(\sqrt{N}) = O(2^{\frac{n}{2}})$ to $O(n)$, where $n$ represents the length of the bit string and also the required number of qubits. This makes our method more feasible and applicable to current NISQ hardware.

\end{itemize}

\subsection{Related Works and Comparison}
There is a related work that try to use variational learning to retrieve the Grover search~\cite{vgs1}. In comparison to their work, which utilizes an oracle-dependent VQE ansatz and a binary value objective function ($f(x \neq \omega) = 1$ and $f(x = \omega) = 0$), with $\omega$ the target bit string, our VQE ansatz is independent of the oracle and the objective function is the expectation value of the Oracle Hamiltonian. The Oracle Hamiltonian has $2n$ possible eigenvalues if $n$ is odd and $2n +1$ possible eigenvalues if $n$ is even. The independence of our VQE ansatz from the oracle allows for a reduction in circuit depth as it is not necessary to encode oracle information in the circuit. Additionally, this approach can improve the robustness with respect to noise. Overall, our method offers a more flexible and generalizable VQE approach that can be applied to a wider range of search tasks, including those in which it is difficult to encode the oracle in the circuit or when the encoded circuit is too deep to be executed on real hardware.

\section{VQE Search}
\label{sec:vqesearch}

We propose a VQE optimization scheme for the search problem to find a target string $\omega$ with length $n$. First, the construction of the Oracle Hamiltonian will be introduced. Second, the details of the VQE optimization process will be described. 

\subsection{Construction of the Oracle Hamiltonian}
In this section, an Oracle Hamiltonian is constructed to find a corresponding target string $\omega$ with length $n$. The ground state of such Hamiltonian will correspond to the target string $\omega$. For a target bit string $\omega \in \{0,1\}^{n}$, an Oracle Hamiltonian is constructed in the following manner: 
\begin{equation}
H_o = \sum_{i = 1}^n h_i \sigma_i^z,
\end{equation}
with the $\omega$-dependent coefficients 
\begin{equation}
h_i =     \left\{ \begin{array}{rcl}
         -1 & \mbox{for} & \omega_i = 1, \\ 
          1 & \mbox{for} & \omega_i = 0,
                \end{array}\right.
\end{equation}
such that for a target bit string state $| \omega \rangle = |x \rangle^{\otimes n}$, where $x \in \{ 0, 1\}$, if we set the corresponding eigenvalue and eigenvectors for $\sigma^z$:
\begin{align}
\sigma^z | 0 \rangle &= - | 0 \rangle \\ 
\sigma^z | 1 \rangle &= | 1 \rangle,
\end{align}

the eigenvalue $-n$ of the Oracle Hamiltonian $H_o$ can be obtained from the eigenvalue equation
\begin{equation}
\label{eq:egv}
    H_o | \omega \rangle = -n | \omega \rangle. 
\end{equation}
Since $h_i \in \{-1, 1\}$ and the eigenvalues of $\sigma^z$, $\lambda_j \in \{-1, 1\}$, in all possible eigenvalues of $H_o$, one can observe that $-n$ is indeed the lowest eigenvalue of $H_o$, with the corresponding ground state $| \omega \rangle$. From the construction of the Oracle Hamiltonian, the task of searching the target bit string $\omega$ become the task of finding the ground state of the Oracle Hamiltonian $H_o$.  


The Oracle Hamiltonian that corresponds to the search problem of the bit string $\omega$ is a 1-local Hamiltonian, which is classified under the P complexity class \cite{hamcplx1}. It should be emphasized that the prior knowledge of $\omega$ is required to construct the Hamiltonian. Furthermore, it is worth noting that problems, such as max $k$-SAT, for which the solution is not known in advance, can be solved using the Grover algorithm \cite{maxsatgrover1}. These problems can be formulated by the $k$-local Hamiltonian \cite{toising1, toising2}, which is categorized under QMA-complete \cite{hamcplx1}.

\subsection{VQE optimization}
In this section, the detail of the VQE optimization will be discussed. From the basic idea of the VQE optimization, the ansatz we chose to implement the idea, to the different optimizers for the simulation and real device experiment.

\begin{figure*}[ht]
\centering
\includegraphics[scale=0.24]{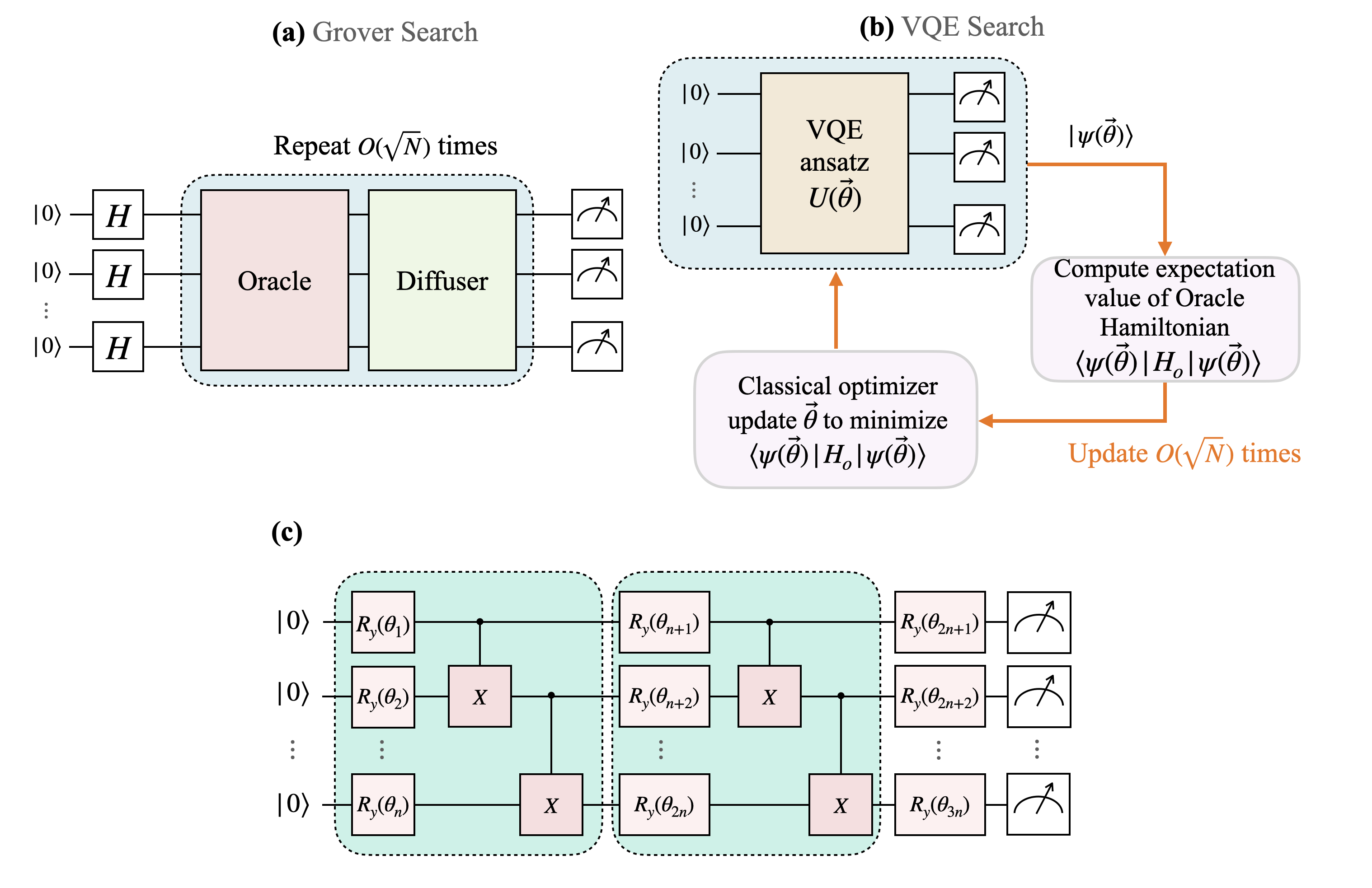}
\caption{(a) Schematic draw of Grover search. (b) Schematic draw of the VQE search. (c) The heuristic VQE ansatz "Real Amplitude" used in this work.}
\label{fig:scheme}
\end{figure*}

The task of optimizing the parameters of VQE requires an objective function, in this work, the objective function is defined as the expectation value of the Oracle Hamiltonian $H_o$ with respect to the parameterized quantum state $| \psi(\vec{\theta}) \rangle$ that provided by the VQE, where $\vec{\theta} = (\theta_1, \theta_2, \hdots, \theta_{3n})$ is the set of the VQE parameter with total $3n$ parameters, with $n$ the number of qubits as well as the length of the target bit string. The VQE ansatz ``Real Amplitude," which has $3n$ tunable parameters, is depicted in Figure~\ref{fig:scheme}(c). It is constructed using 3 layers of $R_y$ rotational gates and 2 layers of CNOT gates with a linear distribution, also known as a linear entangler. As a result of the linear entangler, the required circuit depth grows linearly as the number of qubits increases. 

The goal of the optimization is to find the optimal parameter set $\vec{\theta}_{\text{opt}}$ such that
\begin{equation}
\vec{\theta}_{\text{opt}} = \text{argmin}_{\vec{\theta}} \langle \psi(\vec{\theta}) | H_o | \psi(\vec{\theta}) \rangle,
\end{equation}
where $\langle \psi(\vec{\theta}) | H_o | \psi(\vec{\theta}) \rangle$ is the expectation value of the Oracle Hamiltonian $H_o$. From Eq.~(\ref{eq:egv}), the lowest possible value for $\langle H_o \rangle$ is
\begin{equation}
\min \langle H_o \rangle = \langle \omega | H_o| \omega \rangle = -n,
\end{equation}
where $|\omega \rangle$ is the ground state of $H_o$. Therefore, the optimization task can also be viewed as simulating $|\omega \rangle$ with $| \psi(\vec{\theta}) \rangle$, in order to make $\langle \psi(\vec{\theta}) | H_o | \psi(\vec{\theta}) \rangle$ as close to $-n$ as possible.

In order to make the iterations of the optimization comparable to the $O(\sqrt{N})$ iterations of the Grover search, the VQE optimization also uses $O(\sqrt{N})$ iterations. However, the number of function evaluations required per iteration may vary depending on the choice of classical optimizer. For example, the SPSA optimizer requires approximately 2.5 function evaluations per iteration, while the COBYLA optimizer requires only 1 function evaluation per iteration. Each evaluation of the objective function requires the execution of the VQE circuit, so it is important to carefully consider the total number of circuit executions when implementing on real hardware in order to conserve quantum computational resources. Therefore, in the following sections, the SPSA optimizer is used in the simulation results, and the COBYLA optimizer is used in the real hardware results.

In Fig.~\ref{fig:scheme}(a), the basic scheme flow of the Grover search has been shown, as a comparison, in Fig.~\ref{fig:scheme}(b), the flow of the VQE search is also illustrated. From the comparison above, it can be seen that in the VQE search, the role of the Oracle circuit, which includes the information of the target bit string in the Grover search, becomes the Oracle Hamiltonian and is independent of the circuit. For the task of tuning the target state, the Grover search uses the Diffuser operator or Diffuser circuit, while the VQE search uses the VQE ansatz circuit to perform a similar function. Finally, in the Grover search, the role of implementing the iterative method, which involves the repetitive use of the Oracle-Diffuser combination, becomes the classical optimizer iterations in the VQE search. This substitution reduces the required quantum circuit depth from $O(\sqrt{N}) = O(2^{\frac{n}{2}})$ to $O(n)$, which is a reduction in the $\log$ level.

\section{Results}

\subsection{Quantum Search on ideal and noisy simulators.}
The proposed VQE search method was tested on both ideal (noise-free) and noisy simulators to demonstrate the effect of noise and the superiority of the proposed method over the Grover search in the presence of noise. In this study, we considered the task of finding a bit string $\omega$ of length $n$ among $N = 2^n$ possible bit strings. The success probability was defined as the number of times the target bit string was measured divided by the total number of measurement shots.
As shown in Fig.~\ref{fig:main_res}(a), the label ``ideal Grover" refers to the Grover search implemented on the ideal simulator, and the label ``noisy Grover" refers to the Grover search implemented on the noisy simulator. For each bar representing the results implemented on the simulator, 10 random target bit strings were tested. We can observe that the presence of noise is critical to the success probability of the Grover search, and when $n \ge 4$, the result of the "noisy Grover" is essentially at the same level as that of random guessing, whose success probability is $\frac{1}{N} = \frac{1}{2^n}$. In contrast, the label ``ideal VQE" refers to the VQE search proposed in this work implemented on the ideal simulator, and the label ``noisy VQE" refers to the VQE search implemented on the noisy simulator. Although the success probabilities of the "noisy VQE" are not as high as those of the "ideal VQE", we can observe that the results are still usable, as the success probability is around $0.6$ when $n = 9$. 

For the VQE search, we need to optimize the expectation value of an Oracle Hamiltonian $\langle H_o \rangle$. In this design, the task of finding the target string becomes a minimization task, where the corresponding state with the lowest expectation value corresponds to our target bit string state (the details of which will be described in Sec.~\ref{sec:vqesearch}). For the ``ideal VQE" and ``noisy VQE" labels in Fig.\ref{fig:main_res}(a), we use $10\lfloor \sqrt{N} \rfloor$ iterations during the optimization process with the SPSA optimizer. In Fig.\ref{fig:main_res}(b), we present the expectation value of the Oracle Hamiltonian $\langle H_o \rangle$ during the VQE optimization process with $10 \lfloor \sqrt{N} \rfloor$ iterations for $n \in \{3,5,7\}$. The label ``Ideal" refers to the ideal simulation and the label ``Noisy" refers to the noisy simulation. Note that for the SPSA optimizer we used in the simulator results, each optimization iteration costs an average of 2.5 function evaluations (nfev), which is the horizontal axis of Fig.~\ref{fig:main_res}(b).

\subsection{Quantum Search on NISQ quantum hardware up to 9 qubits.}
\begin{figure*}[ht]
\centering
\includegraphics[scale = 0.32]{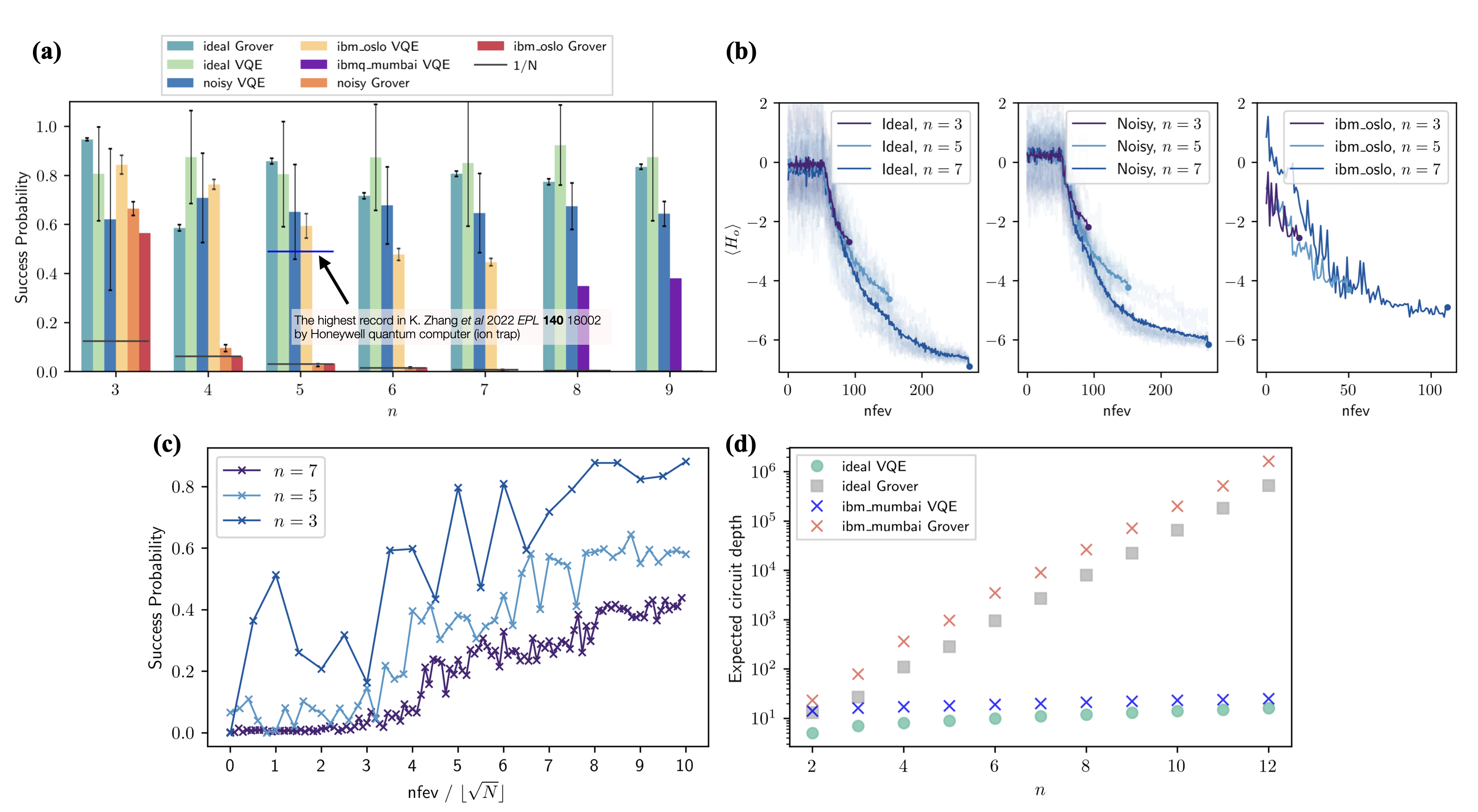}
\caption{(a) The success probability with different number of qubits $n$ on several different computational backends. (b) The expectation value of the Oracle Hamiltonian $\langle H_o \rangle$ during the optimization process of the VQE search. The term "nfev" means number of the function evaluation. (c) The success probability with respect to the nfev in the unit of $\lfloor \sqrt{N} \rfloor$. (d) Expected circuit depth with different number of qubits $n$. }
\label{fig:main_res}
\end{figure*}

In order to evaluate the performance of Grover search and VQE search on a real quantum computer within a reasonable time frame, we conducted experiments with a limited number of target bit strings on the IBMQ platform. For Grover search, we tested 1 random target bit string for each $n$ up to $n = 6$ on the 7-qubit ibm\_oslo device, and the results are depicted in Fig.~\ref{fig:main_res}(a) with the label ``ibm\_oslo Grover". These results are consistent with those obtained from noisy simulations, and show that the success probability of Grover search decreases to the level of random guessing for $n \ge 4$. For VQE search, we tested 2 random target bit strings for $n \le 7$ on the 7-qubit ibm\_oslo device, and 1 random target bit string for $n \ge 8$ on the 27-qubit ibmq\_mumbai device. In each case, we conducted 1 trial per target bit string using the COBYLA optimizer for VQE optimization, with a maximum of $\max (10 \lfloor \sqrt{N} \rfloor, 200)$ iterations. Note that the COBYLA optimizer requires 1 function evaluation per iteration, which helps to keep the number of circuit executions under control when working with real hardware. 

In Fig.~\ref{fig:main_res}(a), the results of the VQE search obtained from real devices are shown in the labels ``ibm\_oslo VQE" and ``ibm\_mumbai VQE". Although the success probability is not as high as that from the noisy simulation, our method still retrieves a usable success probability of approximately 0.4 when $n = 9$. According to K. Zhang's work~\cite{nisqgrover1} published in 2022, the highest success probability of a Grover search implemented on a real quantum computer to date is 0.49 when $n = 5$, and we also indicate their result with a blue line in Fig.~\ref{fig:main_res}(a). Our success probability result is not only higher than the highest record of quantum search on real hardware when $n=5$, but it is also noteworthy that the architecture of the quantum computer used in their work is an ion-trap based chip produced by Honeywell, which is generally considered to be more reliable and less noisy compared to the superconducting architecture we used. In Fig.~\ref{fig:main_res}(b), the expectation value $\langle H_o \rangle$ during the optimization process for $n \in \{3,5,7\}$ on the 7-qubit ibm\_oslo is shown with label ``ibm\_oslo", with $10\lfloor \sqrt{N} \rfloor$ iterations of COBYLA optimization. It is worth noting that in this case the number of function evaluations (nfev) is lower compared to the simulation case, due to the choice of the optimizer.

\subsubsection{Effect of different optimization iterations.}
The original implementation of Grover search requires $O(\sqrt{N})$ iterations to perform the search, while the VQE search can also use $O(\sqrt{N})$ iterations for optimization. In the Fig.~\ref{fig:main_res}(a), the results implemented on the real hardware use $\max (10 \lfloor \sqrt{N} \rfloor, 200)$ iterations. In order to demonstrate the relation of the success probability and the optimization iterations in the real hardware, we present the success probability with respect to the nfev in the unit of $\lfloor \sqrt{N} \rfloor$ in Fig.~\ref{fig:main_res}(c) for $n \in \{3,5,7\}$ obtained from ibm\_oslo.  It can be observed that as $n$ increases, a higher number of function evaluations (nfev) is required to obtain the same level of success probability. Specifically, for $n=7$, it requires approximately $4 \lfloor \sqrt{N} \rfloor$ nfev to obtain a success probability significantly larger than $1/N$. This trend highlights the importance of the number of nfev in order to achieve a desired level of success probability, particularly for larger values of $n$.

\subsubsection{Circuit depth reduction of VQE Search.}
As shown in the previous sections, the success probability of both the noisy simulation and the real hardware significantly differs between Grover search and VQE search, with VQE search outperforming Grover search. The primary reason for this difference in performance is the variation in the circuit depth required by the two search methods. Figure~\ref{fig:main_res}(d) illustrates the expected circuit depth of both Grover search and VQE search on the ideal simulator and ibm\_mumbai. It is worth noting that the quantum circuit implemented on the real hardware requires additional gate operations for SWAP gates due to the connectivity between qubits, as well as for decomposing the gate operations due to the limited set of supported gate operations. These factors contribute to the difference in circuit depth between the ideal simulator and the real hardware implementation. The design of Grover search involves performing all $O(\sqrt{N})$ iterations on the same quantum circuit, which grows exponentially with $n$. As a result, the resulting circuit can be extremely deep and difficult to implement on real hardware due to noise. In contrast, VQE search only requires the construction of the ansatz circuit, which, for the case of ansatzes using linear entanglers, grows linearly with $n$ (as described in more detail in Sec.~\ref{sec:vqesearch}). This makes VQE search more practical to implement on real hardware compared to Grover search, and also emphasizes the importance of considering the circuit depth in the design and implementation of quantum algorithms, as it can significantly affect their performance.

\section{Discussion and Conclusion}


By choosing to replace the $O(\sqrt{N})$ iterations of Grover search on the same quantum circuit with iterations on a classical optimizer and an Oracle Hamiltonian, the proposed VQE search can improve the success probability of the search task on real hardware. We demonstrated that the VQE search method can be performed by constructing a corresponding Oracle Hamiltonian, which maps a quantum search task to an expectation value minimization task, making a search problem practical on NISQ hardware. 

In this work, we utilize the ``Real Amplitude" heuristic VQE ansatz, which is constructed using $R_y$ rotational gates and linearly distributed CNOT gates, as depicted in Fig.~\ref{fig:scheme}(c). While it is possible that other VQE ansatzes may perform better on the VQE optimization problem, for example, to consider the connectivity between the qubits in the actual quantum computer and design a hardware-efficient circuit for the ansatz, we plan to investigate this topic as future work and explore the possibility of finding a more efficient and effective ansatz through a VQE architecture search algorithm.
  
The circuit depth is critical for implementing a quantum algorithm on real hardware. In this work, the linear entangler used in the VQE ansatz in the search task causes the depth of the quantum circuit to grow linearly with $n$. There are also other choices of entangler available, for example, using a fully entangler will result in a circuit depth of $O(\text{poly}(n))$, as $\frac{n(n-1)}{2}$ CNOT gates are required to create a fully entangler structure. It is important to carefully consider the choice of entangler and its impact on the circuit depth in the design and implementation of quantum algorithms, as it can significantly affect their performance on real hardware.

In this study, the number of iterations used in the VQE search is set to $\text{max}(10 \lfloor \sqrt{N} \rfloor, 200)$. Recall that the Grover search achieves a quadratic speedup by going from $O(N)$ to $O(\sqrt{N})$. To make the VQE search more efficient than an exhaustive search with $O(N)$, the condition $10 < \sqrt{N}$ or $10 < 2^{\frac{n}{2}}$ must be satisfied. For integer $n$, we see that $n \ge 7$ satisfies this condition. However, as shown in Fig.~\ref{fig:main_res}(c), even fewer iterations than $10\sqrt{N}$ can still result in success probabilities that are higher than random guessing.

For now, there is no theoretical evidence to suggest that a hybrid quantum-classical optimization, such as the VQE search, can outperform a pure quantum Grover search. However, our experimental results demonstrate that the VQE search is more effective on NISQ hardware than the Grover search. In the future, a fault-tolerant and more powerful quantum computer is expected to improve the performance of the VQE search by enabling us to search for longer bit strings using deeper quantum circuits with higher success probability. While these potential improvements are also applicable to the Grover search, the short sequence circuit nature of the VQE search still makes it more scalable compared to the Grover search. 

\section*{Acknowledgment}
C.Y.L. thanks IBM Quantum Hub at NTU for providing computational resources and accesses for conducting the gate-based quantum computer demonstrations, and
thanks Min-Hsiu Hsieh from Hon Hai Quantum Computing Research Center for valuable discussions.

\end{document}